\documentclass[a4paper,12pt]{article}
\usepackage{graphicx}	
\usepackage{amsmath}	
\usepackage{amssymb}	
\usepackage{subfigure}
\usepackage{sidecap}
\usepackage{tikz}
\usepackage{multirow}
\usepackage{authblk}
\usepackage{txfonts}
\usepackage{newtxtext,newtxmath}
\date{}
\providecommand{\keywords}[1]{\textbf{\textit{Keywords:}} #1}

\begin{document} 
\title{\textbf{Gravitational collapse of compact stars in $ f(R) = \xi R^4 $ gravity}}

\author{Jay Solanki}
\affil{Sardar Vallabhbhai National Institute of Technology,\\ Surat - 395007, Gujarat, India\\ E-mail: jay565109@gmail.com }

\maketitle

\begin{abstract}
\noindent Model of gravitational collapse of anisotropic compact stars in a new theory of $ f(R) $ gravity has been developed. The author considers the modified gravity model of $ f(R) = \xi R^4 $ to investigate a physically acceptable model of gravitational collapse of anisotropic compact stars. First, the author presents a brief review of the development of field equations of gravitational collapse in $ f(R) $ gravity for a particular interior metric for compact stars. Then analytical solutions for various physical quantities of collapsing anisotropic compact stars in $ \xi R^4 $ gravity have been developed. By analyzing plots of various physical parameters and conditions, it is shown that the model is physically acceptable for describing the gravitational collapse of anisotropic compact stars in $ f(R) = \xi R^4 $ gravity. 
\\

\noindent \keywords{$ f(R) $ gravity, anisotropic stars, gravitational collapse}

\end{abstract}

\section{Introduction}
\label{sec1}
General relativity is a very successful theory for describing various astronomical and cosmological phenomena such as the perihelion of Mercury, compact stars such as white dwarfs and neutron stars, the gravitational collapse of compact stars, and the formation of black holes. However, many astronomically observed phenomena such as late-time acceleration of the universe and galaxy rotation curves can not be explained in the tenets of general relativity. In order to understand such phenomena, concepts of dark energy and dark matter have been introduced.\cite{COPELAND_2006,2011,ARUN2017166,10.2307/24112122} However, as an alternative solution, modification in general relativity is being used to understand such phenomena.\cite{Nojiri:2010wj,Nojiri:2017ncd,Nojiri:2003ft,maartens_durrer_2010,Capozziello_2011,RevModPhys.82.451,doi:10.1142/S0219887807001928} One of the popular modification of general relativity is known as $ f(R) $ theory of modified gravity.\cite{RevModPhys.82.451,doi:10.1142/S0219887820500164,doi:10.1142/S0219887815300032,doi:10.1142/S0219887814600068,doi:10.1142/S0219887815500401, Sebastiani_2015} In $ f(R) $ gravity Einstein-Hilbert action is modified by introducing analytical function of curvature scalar $ f(R) $. Thus, field equations of  $ f(R) $ gravity can be found by varying modified Einstein-Hilbert action. 
\par By choosing particular forms of $ f(R) $ and using modified field equations of $ f(R) $ theories, one can explain various astronomical phenomena that can not be explained by general relativity without invoking dark matter and dark energy concepts. For example the function $ f(R) = \xi R^2 $ has been proposed as inflation model as well as dark matter model.\cite{Starobinsky:1980te,PhysRevLett.102.141301,Cembranos_2011} The function $ f(R) = \xi \frac{1}{R} $ has been proposed as a dark energy model.\cite{PhysRevD.70.043528} Also the function $ f(R) = \xi R^4 $ is proposed as a new candidate for explaining topological inflation.\cite{Kaneda_2010} It has been confirmed that this model describes the existence of slow-roll inflation. Thus, it is beneficial to understand various astronomical phenomena such as inflation, formation of compact stars, and its gravitational collapse governed by modified $f(R)$ gravity utilizing various $ f(R) $ models. Many authors have studied formation of compact stars and mass-radius relation for particularly neutron stars in the paper \cite{PhysRevD.93.023501}. They used particular forms of Lagrangians $f(R) = R + \alpha R^2(1 + \gamma R)$ and $f(R) = R^{1+\epsilon}$. Also the authors have studied Extreme neutron stars from Extended Theories of Gravity \cite{Astashenok_2015}. They discussed the neutron stars with the strong magnetic mean field generated by magnetic properties of particles. They described the phenomena by assuming a model with three meson fields and baryons octet. Huge efforts also have been made for Extended Gravity description for the $GW190814$ Supermassive Neutron Star and in studying Causal limit of neutron star maximum mass in $f(R)$ gravity in view of $GW190814$\cite{2020,2021}.
\par The gravitational collapse of compact stars has been extensively studied by various authors. Considering the assumptions of spherically symmetric and pressureless collapsing fluid, exact solutions of gravitational collapse are found in terms of the cycloid parametric equation in this reference\cite{Weinberg:1972kfs}. Various authors also studied the gravitational collapse of compact stars in the framework of $ f(R) $ modified gravity. In \cite{article1}, the authors studied gravitational collapse for various physically important models of $ f(R) $ gravity. A comparative study of gravitational collapse in General Relativity and in R2-gravity is given in this reference\cite{Astashenok2018GravitationalCI}. \cite{article6} studied gravitational collapse for charged adiabatic LTB configuration. Also, \cite{PhysRevD.90.084011} have studied gravitational collapse in the Starobinski model and found an exact solution for an inhomogeneous collapsing star.  Recently, authors have investigated three different $ f(R) $ gravity models to study the collapsing phenomena as well as the nature of central singularity\cite{Jaryal:2021lsu}. Gravitational collapse and formation of black-hole is also studied in other physically important theories by many authors.\cite{Bhatti2020SphericalCW, Cembranos2011KERRNEWMANBH, 2016IJGMM..1330005G, Andrasson2014OnGC, Chakraborty_2020} In \cite{2008GReGr..40.2149P}, the authors revisited radiating gravitational collapse with shear viscosity. In \cite{Chowdhury:2019phb}, the authors studied gravitational collapse for various separable and non-separable forms of interior metric and applied method of $ R $ matching to generate physically acceptable models in $ f(R) $ gravity. It was found that a very simplified form of the separable interior metric is not suitable for describing this problem. Thus, a particular form of separable interior metric admitting homothetic killing vectors and another non-separable form of interior metric can completely describe the collapse in $ f(R) $ gravity. Thus, gravitation collapse in $ f(R) $ gravity is a very active research area.\cite{Kolassis_1988,PhysRevD.88.084015,PhysRevD.90.024017,PhysRevD.90.084011,Rudra_2014}
\par To study the gravitational collapse in the $ f(R) $ gravity framework, it is necessary to formulate field equations of $ f(R) $ gravity first. The field equations of $ f(R) $ gravity can be formulated by varying modified Einstein-Hilbert action. After that, one needs to solve the equations for a particular interior metric. Many authors studied gravitational collapse for various separable and non-separable forms of interior metric. The gravitational collapse in $ f(R) $ gravity for homogeneous Friedmann-Robertson-Walker metric has been extensively studied. However, recently it was found that the assumption of a separable solution of that form or similar form is an oversimplification of the problem.\cite{Chowdhury:2019phb} Thus, another separable form of the metric admitting a homothetic killing vector has been considered to formulate gravitational collapse of compact stars in $ f(R) $ gravity.\cite{Wagh_2001,wagh2002axially} Gravitational collapse for that metric has been studied for $ R^2 $ modified gravity.\cite{Chowdhury:2019phb} As pointed out earlier, it is useful to study such collapse in $ R^4 $ modified gravity, which is a candidate for slow-roll inflation. Thus, for such a model of modified gravity($ f(R) = \xi R^4 $), in this paper, the author investigates the gravitational collapse in $ R^4 $ gravity for interior space-time admitting homothetic killing vectors.
\par The author organizes this paper as follows: In section \ref{sec2}, the author presents field equations of $ f(R) $ gravity. In section \ref{sec3}, the author briefly reviews the field equations of gravitational collapse in $ f(R) $ gravity for interior space-time admitting homothetic killing vectors. The reason for providing a brief review on the derivation of field equations is to remove ambiguity in different solutions circulating for the same interior metric. Also, the author computes and shows non-zero components of Einstein tensor and matter and curvature energy-momentum tensor, which may be useful for some readers. In section \ref{sec4}, the author develops analytical solutions of various physical quantities such as energy density and two pressures for a collapsing star in $ f(R) = \xi R^4 $ gravity. In section \ref{sec5}, the author describes various physical conditions on the model of gravitational collapse to be physically acceptable. The author plots various physical quantities of stars and physical conditions and shows that the particular model of gravitational collapse is physically acceptable for describing the gravitational collapse of anisotropic compact stars in $ f(R) = \xi R^4 $ gravity. In section \ref{sec6}, the author considers the physical analysis of the model. In section \ref{sec7} the author investigates the physical properties of the stars undergoing gravitational collapse. Then author develops junction conditions in section \ref{sec8} to be imposed on gravitationally collapsing stars. Finally, the author concludes the results obtained in this paper in section \ref{sec9}.

\section{Field Equation of $ f(R) $ Gravity}
\label{sec2}
Field equations of $ f(R) $ gravity is formulated by modifying Einstein-Hilbert action of General Relativity. Modification of Einstein-Hilbert action can be done by introducing general function of curvature $ f(R) $ in the Einstein-Hilbert action. Thus, Lagrangian density for modified action becomes $ R + f(R) + \mathcal{L}_{matter} $. Variation of action written in terms of that given Lagrangian density yields following field equation of $ f(R) $ gravity
\begin{equation}
    \label{1}
    G_{\alpha \beta} = \frac{1}{1+f_R}\left( T_{\alpha \beta} + \nabla_{\alpha}\nabla_{\beta}f_R - g_{\alpha \beta}\nabla_{\mu}\nabla^{\mu}f_R + \frac{1}{2}g_{\alpha \beta}(f - R f_R) \right)
\end{equation}
Where $ G_{\alpha \beta} $ is Einstein tensor given by $ G_{\alpha \beta} = R_{\alpha \beta} - \frac{1}{2}g_{\alpha \beta}R $. $ T_{\alpha \beta} $ is the energy momentum tensor. Also $ f_R = \frac{df(R)}{dR} $. Now for anisotropic matter distribution of vanishing shear, we consider following simplified form of energy momentum tensor\cite{de2010f}
\begin{equation}
    \label{2}
    T_{\alpha \beta} = (\rho + p_t)u_{\alpha}u_{\beta} + p_t g_{\alpha \beta} - (p_t - p_r)v_{\alpha}v_{\beta} + 2qu_{(\alpha} v_{{\beta})}
\end{equation}
Where $\rho$, $ p_r $ and $ p_t $ denotes energy density, radial pressure and tangential pressure respectively. $ u_{\alpha} $ and $v_{\alpha}$ denotes the four-velocity of matter and  space-like vector in the radial direction, respectively. Also $ q^{\alpha} = q v^{\alpha} $ is the radial heat flow vector. Now the normalization condition of each vector is given by $ v^{\alpha} v_{\alpha} = 1$, $ u^{\alpha} u_{\alpha} = -1 $ and $ u^{\alpha} q_{\alpha} = 0$
Now equation (\ref{1}) can be written in the following form
\begin{equation}
    \label{3}
    G_{\alpha \beta} = \frac{1}{1+f_R}\left( T^m_{\alpha \beta} + T^c_{\alpha \beta} \right)
\end{equation}
Where $ T^m_{\alpha \beta} $ is the stress energy tensor of matter given by equation (\ref{2}) for anisotropic matter distribution. $T^c_{\alpha \beta}$ is the effective stress energy tensor which has purely geometric origin. In this sense $T^c_{\alpha \beta}$ is the additional source of the curvature in the $ f(R) $ gravity. From equation (\ref{1}), $T^c_{\alpha \beta}$ is given by
\begin{equation}
    \label{4}
    T^c_{\alpha \beta} = \nabla_{\alpha}\nabla_{\beta}f_R - g_{\alpha \beta}\nabla_{\mu}\nabla^{\mu}f_R + \frac{1}{2}g_{\alpha \beta}(f - R f_R)
\end{equation}
\section{Review of Field Equations of Gravitational Collapse in $ f(R) $ Gravity}\label{sec3}
Now in order to study spherically symmetric gravitational collapse of anisotropic compact star, we consider following form of the interior metric of the star\cite{Wagh_2001,wagh2002axially}
\begin{equation}
    \label{5}
    ds^2 = - y(r)^2dt^2 + 2c(t)^2\left(\frac{dy}{dr}\right)^2dr^2 + c(t)^2y(r)^2(d\theta^2 + sin^2\theta d\phi^2)
\end{equation}
Now the author solves equation (\ref{1}) for the interior metric (\ref{5}) to study the gravitational collapse of anisotropic compact star in $ f(R) $ gravity. To solve equation (\ref{1}), the author computes non-zero components of tensors $ G_{\alpha \beta} $, $T^m_{\alpha \beta}$ and $T^c_{\alpha \beta}$. Then the author uses equation (\ref{3}) to find the field equations of gravitational collapse of anisotropic compact star in $ f(R) $ gravity for the interior metric (\ref{5}). 
\par The non-zero components of $ G_{\alpha \beta} $ for the interior metric (\ref{5}) are computed as follow
\begin{equation}
    \label{6}
    G_{00} = \frac{1 + 6 \dot{c}^2}{2c^2}
\end{equation}
\begin{equation}
    \label{7}
    G_{01} = G_{10} = \frac{2 \dot{c} y'}{c y}
\end{equation}
\begin{equation}
    \label{8}
    G_{11} = \frac{y'^2}{y^2}\left(1-2\dot{c}^2-4c\ddot{c}\right)
\end{equation}
\begin{equation}
    \label{9}
    G_{22} = \frac{1}{2} - 2c\ddot{c} - \dot{c}^2
\end{equation}
\begin{equation}
    \label{10}
    G_{33} = (sin^2\theta) G_{22}
\end{equation}

we compute non-zero components of $T^m_{\alpha \beta}$ from equation (\ref{2}) as follow

\begin{equation}
    \label{11}
    T^m_{00} = \rho y(r)^2
\end{equation}
\begin{equation}
    \label{12}
    T^m_{01} = T^m_{10} = -q \sqrt{2}c(t)y(r)y'
\end{equation}
\begin{equation}
    \label{13}
    T^m_{11} = p_r 2c(t)^2y'^2 
\end{equation}
\begin{equation}
    \label{14}
    T^m_{22} = p_t c(t)^2y(r)^2
\end{equation}
\begin{equation}
    \label{15}
    T^m_{33} = (sin^2\theta)T^m_{22}
\end{equation}

Now we compute non-zero components of $T^c_{\alpha \beta}$ from equation (\ref{4}),

\begin{equation}
    \label{16}
    T^c_{00} = \frac{y^2f_R''y' + (2yy'^2-y^2y'')f_R' - 6c\dot{c}\dot{f_R}y'^3}{2c^2y'^3} - \frac{1}{2}y^2(f-Rf_R)
\end{equation}
\begin{equation}
    \label{17}
    T^c_{01} = T^c_{10} = \frac{cy\dot{f_r}' - y\dot{c}f_R' - c\dot{f_R}y'}{cy}
\end{equation}
\begin{equation}
    \label{18}
    T^c_{11} = \frac{4c\dot{c}\dot{f_r}y'^2 + 2c^2\ddot{f_R}y'^2 - 3y f_R' y'}{y^2} + c^2y'^2(f-Rf_R)
\end{equation}
\begin{equation}
    \label{19}
    T^c_{22} = \frac{4c\dot{c}\dot{f_R}y'^3 + 2c^2\ddot{f_R}y'^3 - y^2 f_R'' y' - (2yy'^2 - y^2y'')f_R'}{2y'^3} + \frac{1}{2}c^2y^2(f-Rf_R)
\end{equation}
\begin{equation}
    \label{20}
    T^c_{33} = (sin^2\theta)T^c_{22}
\end{equation}

From equation (\ref{3}) and equations (\ref{6}) to (\ref{20}), the author obtains independent field equations of gravitational collapse in $ f(R) $ gravity for metric (\ref{5}) as follow,

\begin{multline}
    \label{21}
    \left(\frac{1 + 6 \dot{c}^2}{2c^2}\right)(1+f_R) = \rho y^2 - \frac{1}{2}y^2(f-Rf_R) \\+ \frac{y^2f_R''y' + (2yy'^2-y^2y'')f_R' - 6c\dot{c}\dot{f_R}y'^3}{2c^2y'^3} 
\end{multline}

\begin{equation}
    \label{22}
    \left(\frac{2 \dot{c} y'}{c y}\right)(1+f_R) = -q \sqrt{2}c y y' + \frac{cy\dot{f_r}' - y\dot{c}f_R' - c\dot{f_R}y'}{cy}
\end{equation}

\begin{multline}
    \label{23}
    \frac{y'^2}{y^2}\left(1-2\dot{c}^2-4c\ddot{c}\right)(1+f_R) = p_r 2c^2y'^2 +\\ \frac{4c\dot{c}\dot{f_r}y'^2  + 2c^2\ddot{f_R}y'^2 - 3y f_R' y'}{y^2} + c^2y'^2(f-Rf_R)
\end{multline}

\begin{multline}
    \label{24}
    \left(\frac{1}{2} - 2c\ddot{c} - \dot{c}^2\right)(1+f_R) = p_t c^2y^2 + \frac{1}{2}c^2y^2(f-Rf_R) \\+ \frac{4c\dot{c}\dot{f_R}y'^3 + 2c^2\ddot{f_R}y'^3 - y^2 f_R'' y' - (2yy'^2 - y^2y'')f_R'}{2y'^3} 
\end{multline}

Where equations (\ref{21}) to (\ref{24}) are field equations of gravitational collapse in $ f(R) $ gravity for the interior metric (\ref{5}). Also overhead dot indicates differentiation with respect to time t and prime indicates differentiation with respect to radial coordinate r. 

\section{Gravitational Collapse in $ \xi R^4 $ Gravity}
\label{sec4}
In this section, the author solves equations of gravitational collapse (\ref{21}) to (\ref{24}) for the particular model of modified gravity $ f(R)  = \xi R^4 $. The reason for choosing this model is that, as mentioned in the introduction, this model describes the existence of slow-roll inflation. Thus, it is useful to study other astronomical phenomena like a gravitational collapse in such a physically important $ f(R) $ gravity model. 
\par Now in this model of $ f(R) $ gravity, $ f(R) = \xi R^4 $ and $ f_R = 4 \xi R^3 $. Now The Ricci scalar for the interior metric (\ref{5}) is given by

\begin{equation}
    \label{25}
    R = \frac{6\dot{c}^2 + 6c\ddot{c} - 1}{c^2y^2}
\end{equation}
Now in order to find physically acceptable model of collapsing star, it is necessary for Ricci scalar an its derivative to be continuous across the matching hyper-surface. Now for the interior metric (\ref{5}), it is enough to choose $ y(r) = (1-r)^{-m} $, with m being a constant of $ m \ge 1 $, for the continuity of the Ricci scalar and its derivative across the matching hyper-surface.\cite{Chowdhury:2019phb} Now by considering particular form of $ y(r) $ for the continuity of the Ricci scalar and its derivative, its clear from equations (\ref{21}) to (\ref{24}) that we are free to choose the function $ c(t) $. Thus the author choose the function $ c(t) $ such that problem simplifies by considering $ 6\dot{c}^2 + 6c\ddot{c} = 0 $ in equation (\ref{25}). The solution of that equation can be calculated and is turns out to be of the form $ c(t) = \sqrt{1-bt} $, where b is an integral constant. 
\par The solutions  $ y(r) = (1-r)^{-m} $ and $ c(t) = \sqrt{1-bt} $ obeys the conditions of continuity of Ricci scalar and its derivative across the matching hyper-surface. Also in the reference \cite{Chowdhury:2019phb}, it has been shown that such solutions indeed obeys all junction conditions for gravitational collapse governed by the interior metric (\ref{5}) in $ f(R) $ gravity.  
\par Now to find physical quantities of collapsing star, the author choose $m = 1$ for the quantity $ y(r) = (1-r)^{-m} $.Now the author formulates analytical solutions for physical quantities of collapsing star with the general function $ c(t) = \sqrt{1-bt} $. Thus two functions of the model is given by

\begin{equation}
    \label{26}
    y(r) = \frac{1}{1-r} \quad\mathrm{and}\quad c(t) = \sqrt{1-bt}
\end{equation}

Where b is integral constant, the value of b will be found such that the model of gravitational collapse becomes physically acceptable in the $ f(R) = \xi R^4 $ gravity.
\par Ricci scalar (\ref{25}) can be found from equation (\ref{26}) as

\begin{equation}
    \label{27}
    R = -\frac{(1-r)^2}{1-bt}
\end{equation}

By substituting the values of equations (\ref{26}) and (\ref{27}), for the model $ f(R) = \xi R^4 $ in the field equations (\ref{21}) to (\ref{24}), the author finds the following analytical solutions of the physical quantities of the collapsing star. 

\begin{multline}
\label{28}
    \rho = \frac{(1-r)^2}{4(1-bt)^5}((2+3b^2) - (8b+9b^3)t + (12b^2+9b^4)t^2 - (8b^3+3b^5)t^3 + 2b^4t^4 \\+ \xi (1-r)^6 ((226+60b^2) - 226bt))
\end{multline}

\begin{equation}
\label{29}
    q = \frac{b(1-r)^2}{\sqrt{2}(1-bt)^{\frac{9}{2}}}(1 - 3bt + 3b^2t^2 + b^3t^3  + 92\xi(1-r)^6)
\end{equation}

\begin{multline}
\label{30}
    p_r = \frac{(1-r)^2}{4(1-bt)^5}((2+b^2) - (8b+3b^3)t + (12b^2+3b^4)t^2 - (8b^3+b^5)t^3\\ + 2b^4t^4 + \xi (1-r)^6 ((142+140b^2) - 142bt))
\end{multline}

\begin{multline}
\label{31}
    p_t = \frac{(1-r)^2}{4(1-bt)^5}((2+b^2) - (8b+3b^3)t + (12b^2+3b^4)t^2 - (8b^3+b^5)t^3\\ + 2b^4t^4 - \xi (1-r)^6 ((242-140b^2) - 242bt))
\end{multline}

Where equations (\ref{28}) to (\ref{31}) are analytical solutions for physical quantities of collapsing star in $ f(R) = \xi R^4 $ gravity for this model. 

\section{Physical Conditions for the Model of Gravitational Collapse}
\label{sec5}
In this section, the author applies various physical conditions to the model of gravitational collapse for it to be physically acceptable. The author considers shear-less fluid and thus applies the following physical conditions to the model. 

\par The basic physical conditions on the model of any physically acceptable stars are that, physical quantities of the stars like energy density $ \rho $, heat flux density $ q $, radial pressure $ p_r $ and tangential pressure $ p_t $ must be of positive values. 

\begin{equation}
    \label{32}
    \rho > 0 \quad\mathrm{and}\quad q > 0
\end{equation}

\begin{equation}
    \label{33}
     p_r > 0 \quad\mathrm{and}\quad p_t > 0
\end{equation}

\par The next physical conditions are that eigenvalues of the energy-momentum tensor must be real for a physically acceptable model of any compact star. The eigenvalues of the energy-momentum tensor can be found by equation $ |T_{\alpha \beta} - \lambda g_{\alpha \beta}| = 0 $. For shear-less fluid, these conditions can be written as

\begin{equation}
    \label{34}
    |\rho + p_r| - 2|q| \ge 0
\end{equation}
\begin{equation}
    \label{35}
    \rho - p_r + 2p_t + \Delta \ge 0
\end{equation}
\\
where, $\Delta = \sqrt{(\rho + p_r)^2 - 4q^2} $.

For a physically acceptable model of gravitational collapse, it must also obey the weak energy condition. Weak energy condition implies that if $ \lambda_0 $ denotes the eigenvalue corresponding to the time-like eigenvector, then it must follow $ -\lambda_0 \ge 0 $. Thus the weak energy condition for shear-less fluid becomes
\begin{equation}
\label{36}
    \rho - p_r + \Delta \ge 0
\end{equation}

Finally physically acceptable model can be found by imposing last strong energy condition. strong energy condition implies that $ \lambda_0 + \sum_{i} \lambda_i \ge 0 $. Which implies the condition
\begin{equation}
\label{37}
    2p_t + \Delta \ge 0
\end{equation}
\par Thus, any physically acceptable model of gravitational collapse must obey all the physical conditions from (\ref{32}) to (\ref{37}). 

\section{Physical analysis of the model}
\label{sec6}
\begin{figure*}[ht!]
\begin{center}
\includegraphics[width = .75\textwidth]{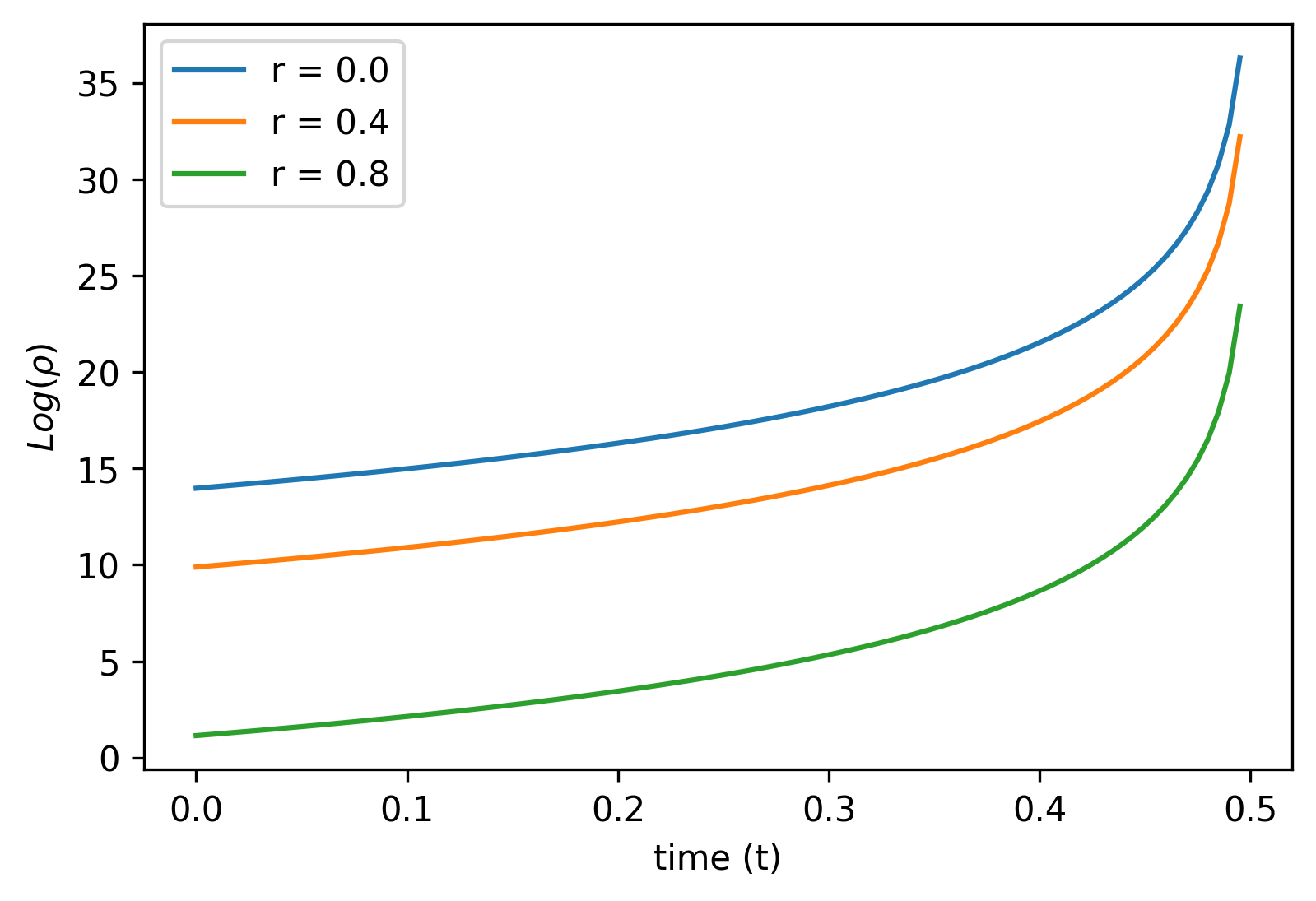}
\caption{ Variation of $Log(\rho)$ against time $ (t) $. }
\end{center}
\label{fig1}
\end{figure*}

\begin{figure*}[ht!]
\begin{center}
\includegraphics[width = .75\textwidth]{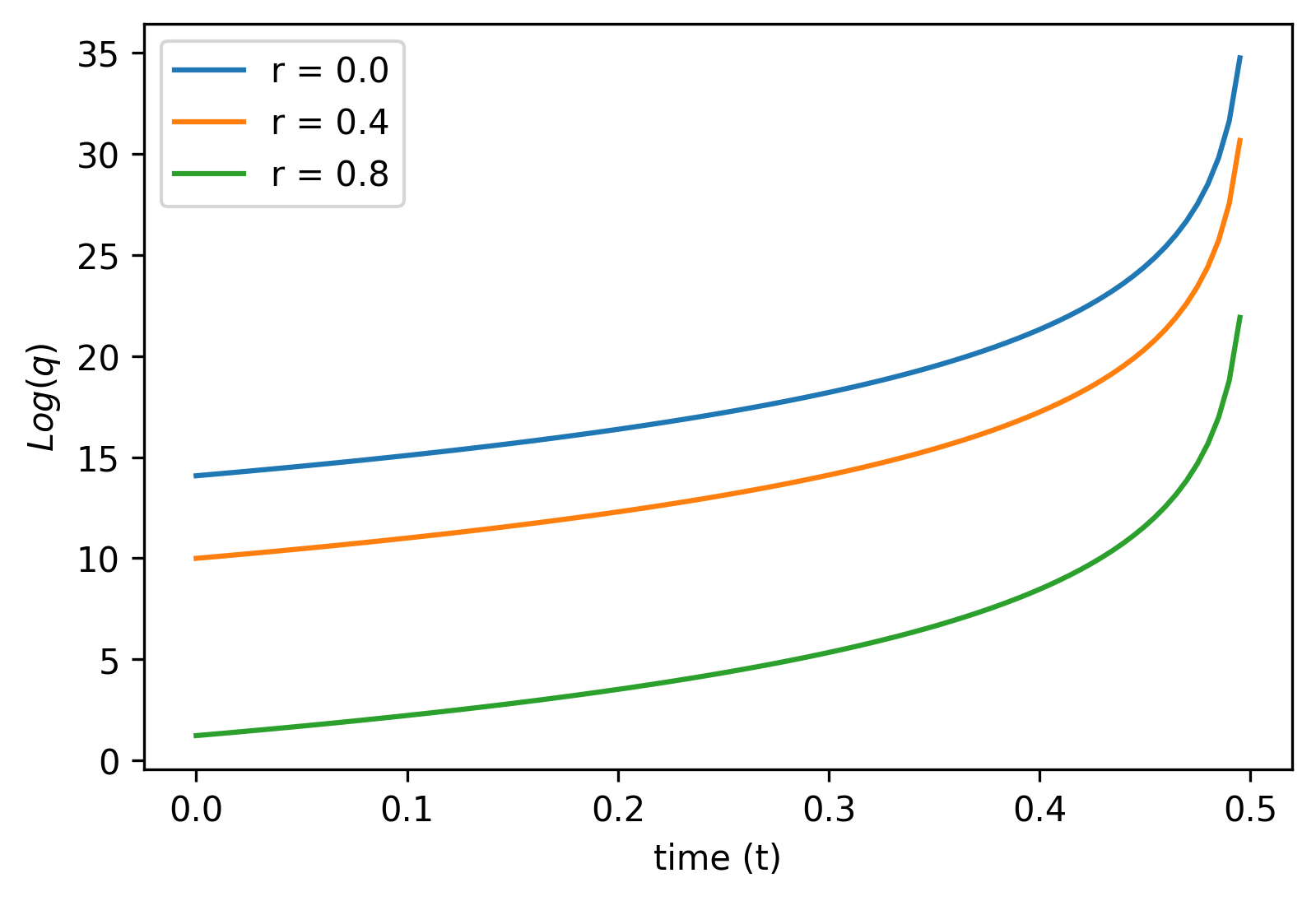}
\caption{ Variation of $ Log(q) $ against time $ (t) $. }
\end{center}
\label{fig2}
\end{figure*}

\begin{figure*}[ht!]
\begin{center}
\includegraphics[width = .75\textwidth]{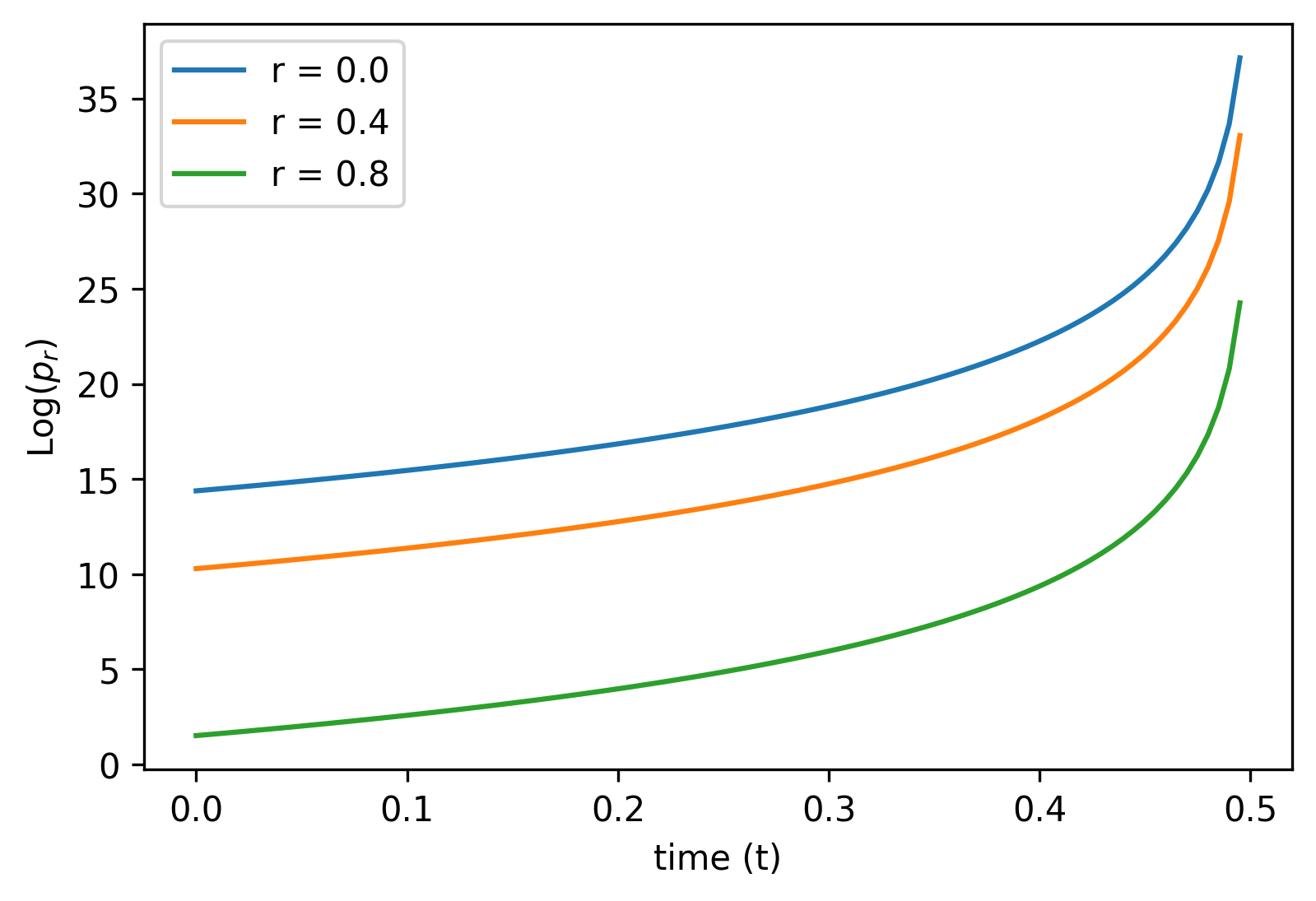}
\caption{ Variation of $Log(p_r)$ against time $ (t) $. }
\end{center}
\label{fig3}
\end{figure*}

\begin{figure*}[ht!]
\begin{center}
\includegraphics[width = .75\textwidth]{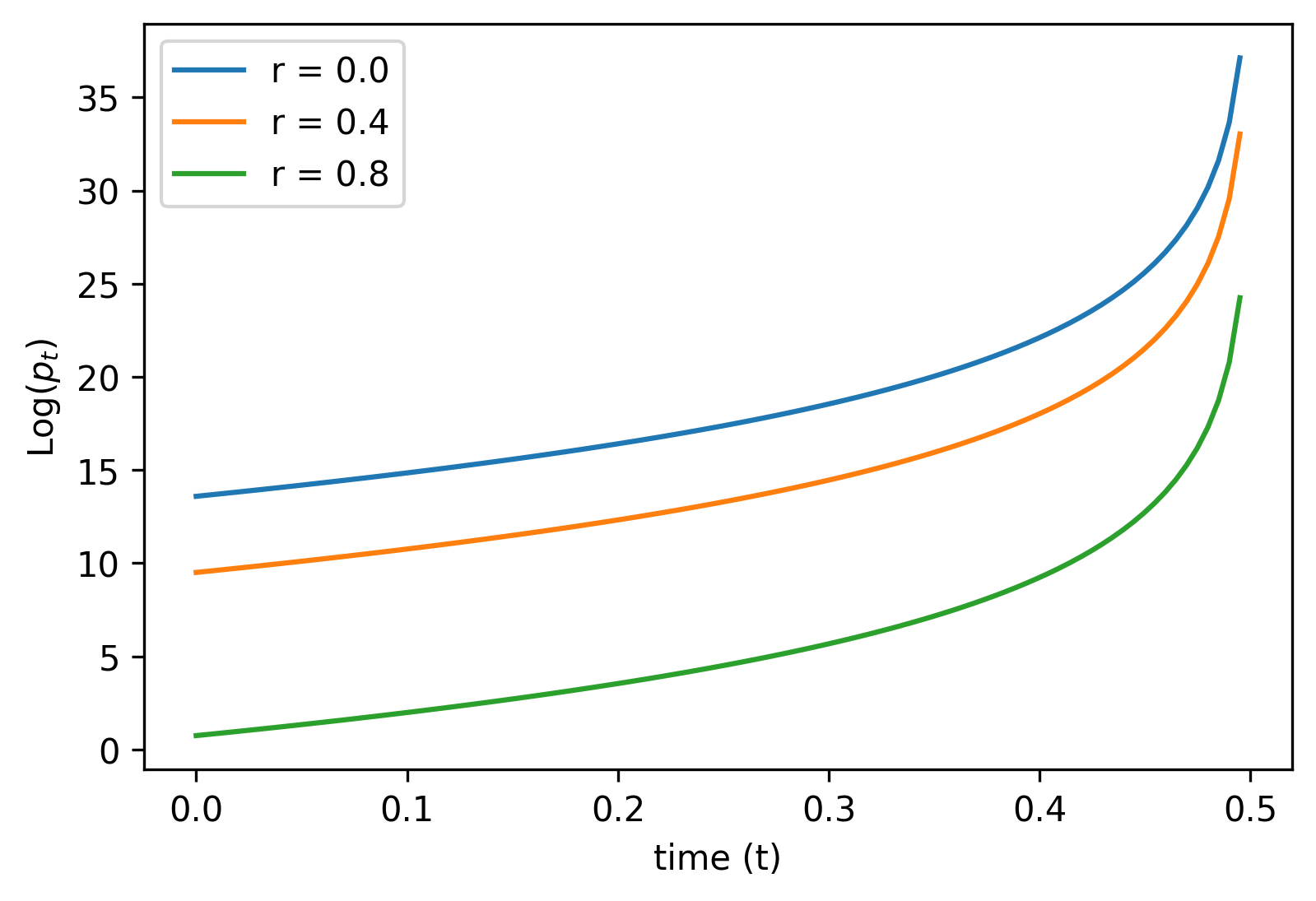}
\caption{ Variation of $Log(p_t)$ against time $ (t) $. }
\end{center}
\label{fig4}
\end{figure*}

\begin{figure*}[ht!]
\begin{center}
\includegraphics[width = .75\textwidth]{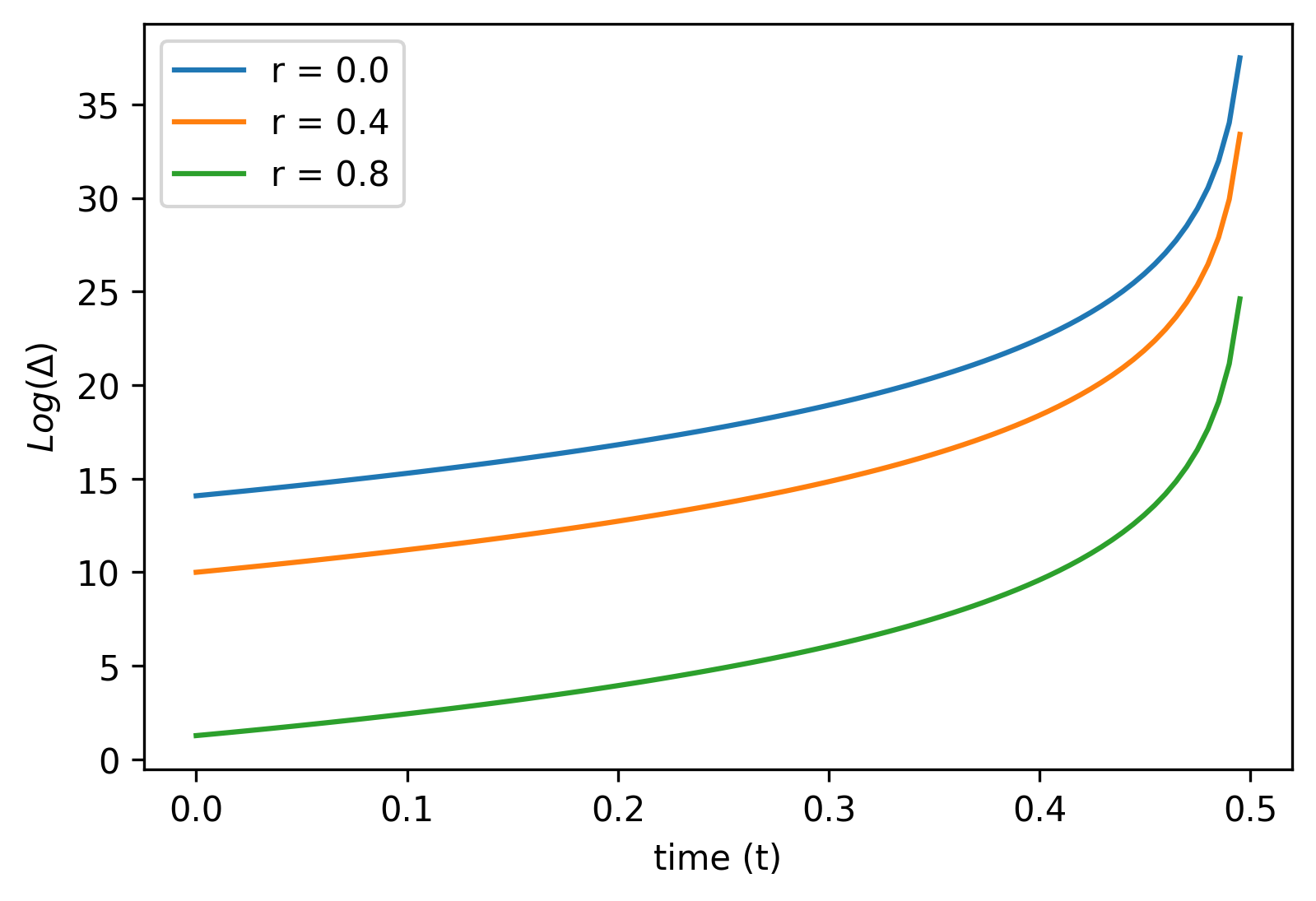}
\caption{ Variation of $Log(\Delta)$ against time $ (t) $. }
\end{center}
\label{fig5}
\end{figure*}

\begin{figure*}[ht!]
\begin{center}
\includegraphics[width = .75\textwidth]{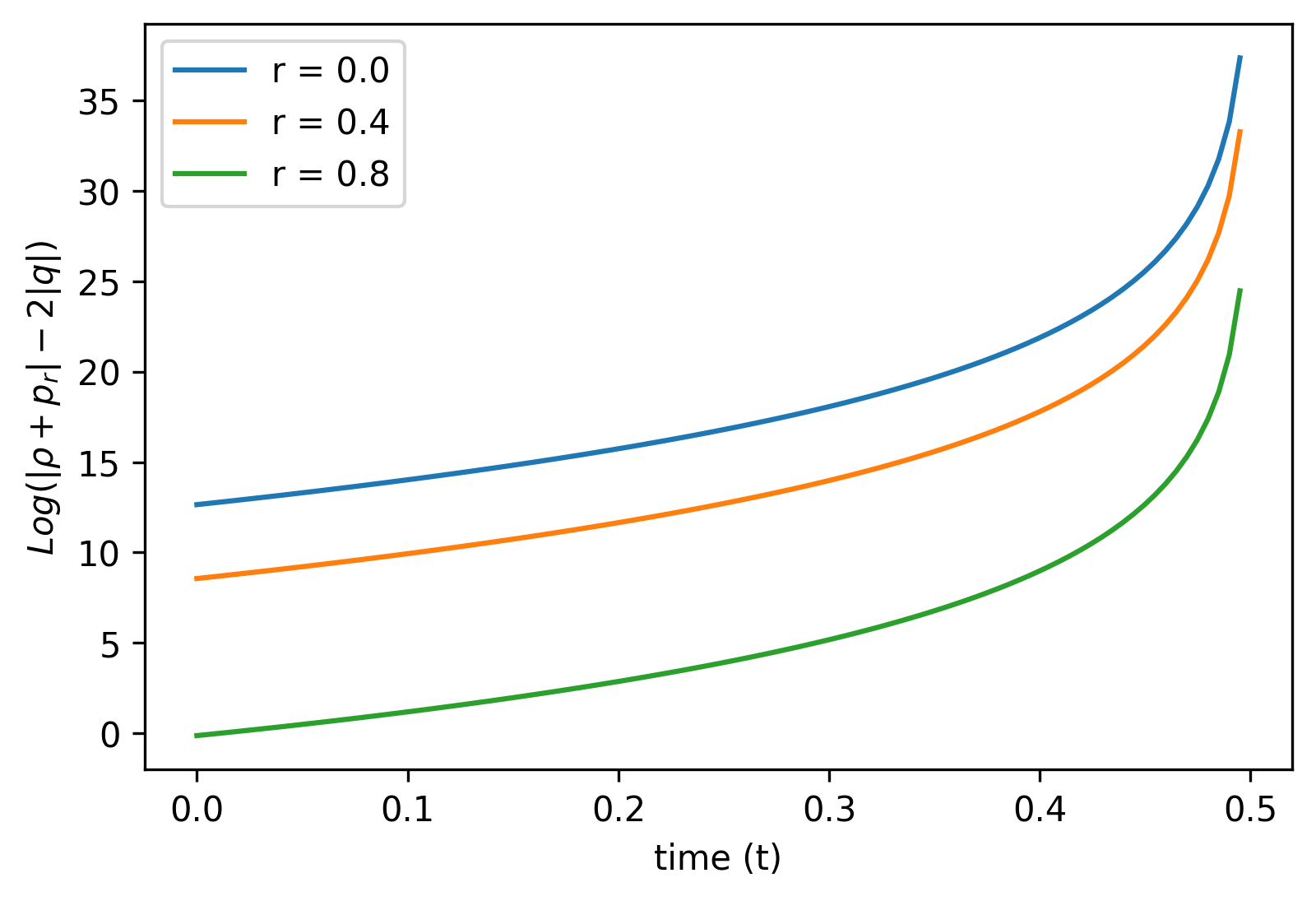}
\caption{ Variation of $ Log(|\rho + p_r| - 2|q|) $ against time $ (t) $. }
\end{center}
\label{fig6}
\end{figure*}

\begin{figure*}[ht!]
\begin{center}
\includegraphics[width = .75\textwidth]{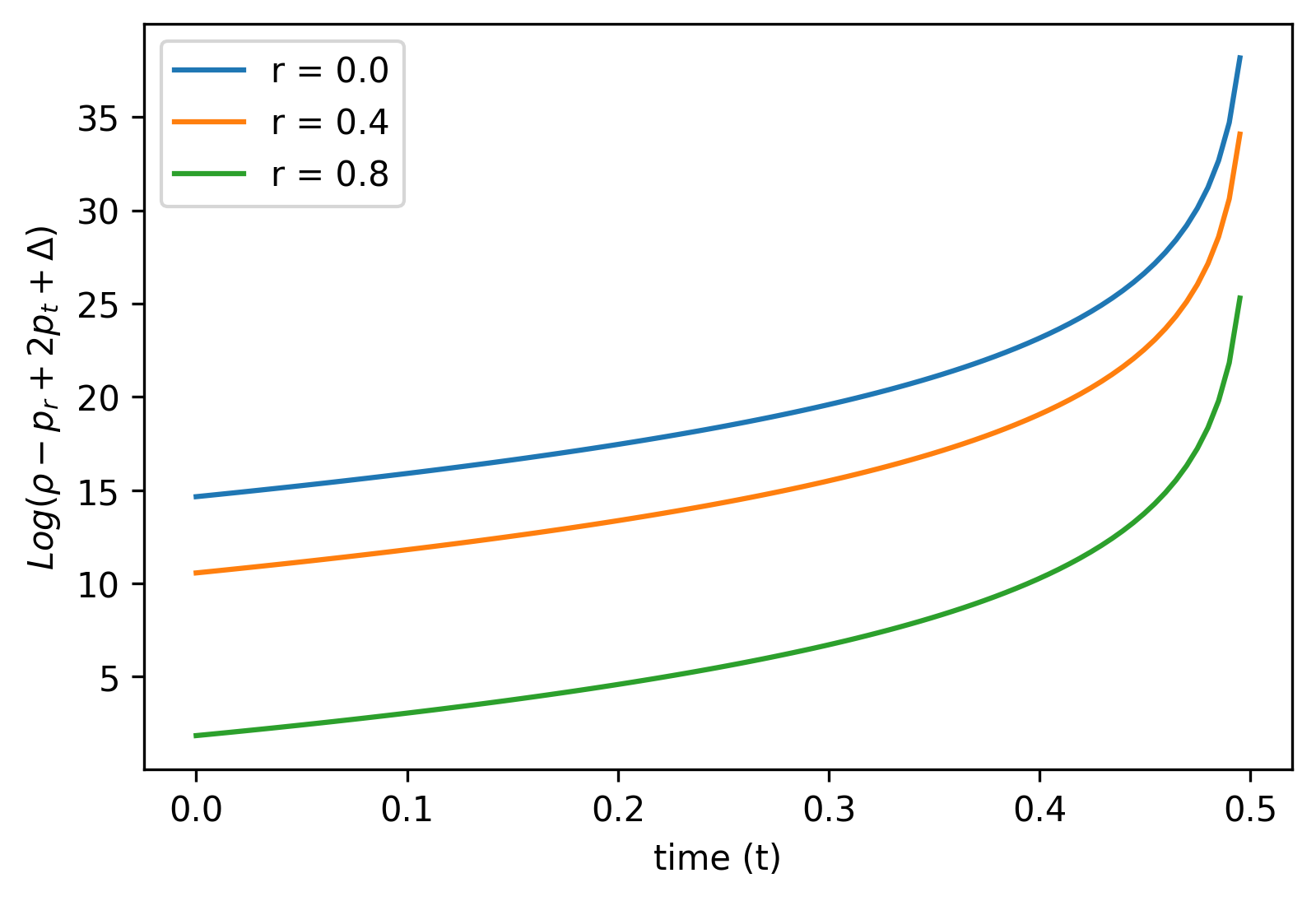}
\caption{ Variation of $ Log(\rho - p_r + 2p_t + \Delta) $ against time $ (t) $. }
\end{center}
\label{fig7}
\end{figure*}

\begin{figure*}[ht!]
\begin{center}
\includegraphics[width = .75\textwidth]{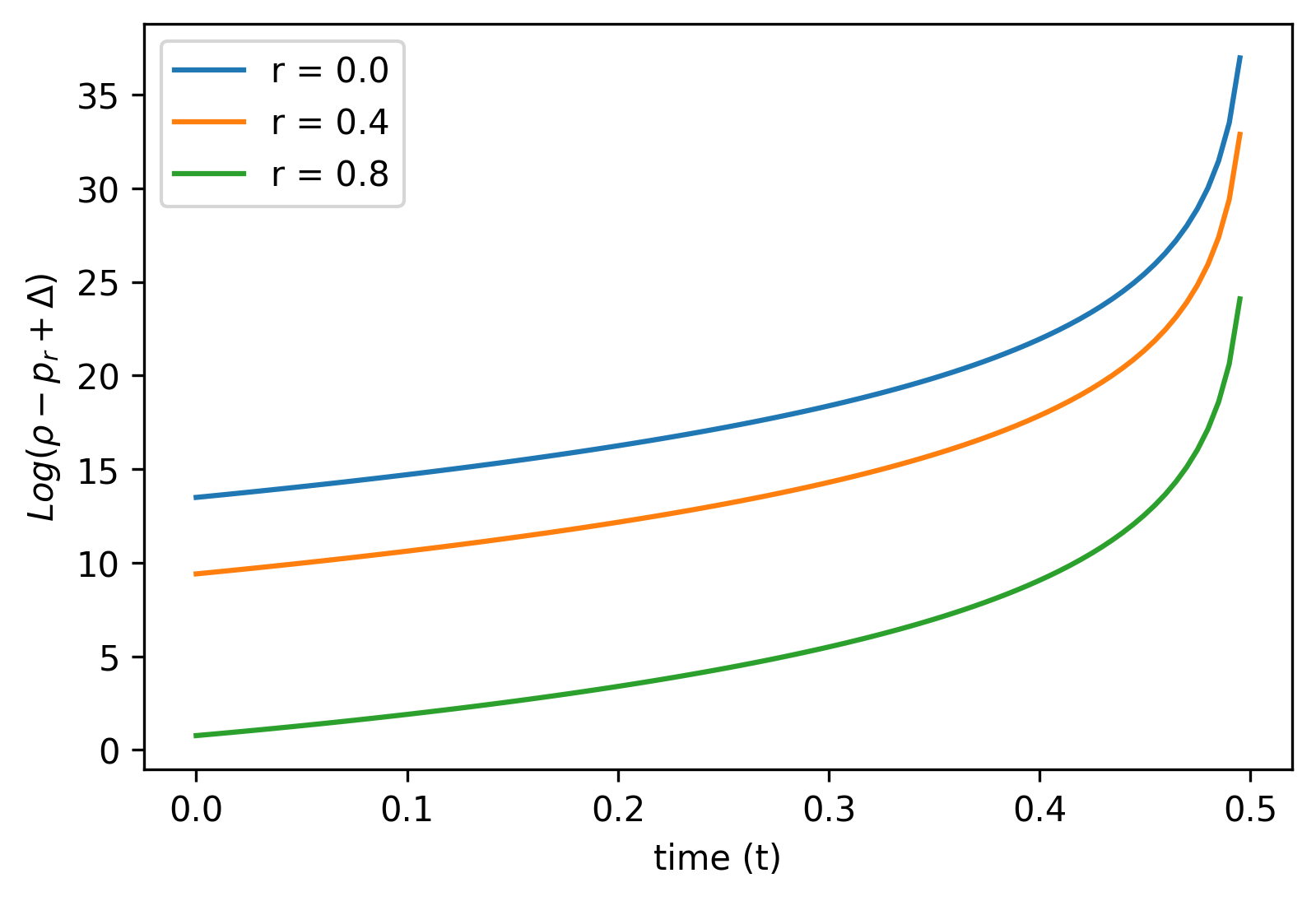}
\caption{Variation of $  Log(\rho - p_r + \Delta) $ against time $ (t) $. }
\end{center}
\label{fig8}
\end{figure*}

\begin{figure*}[ht!]
\begin{center}
\includegraphics[width = .75\textwidth]{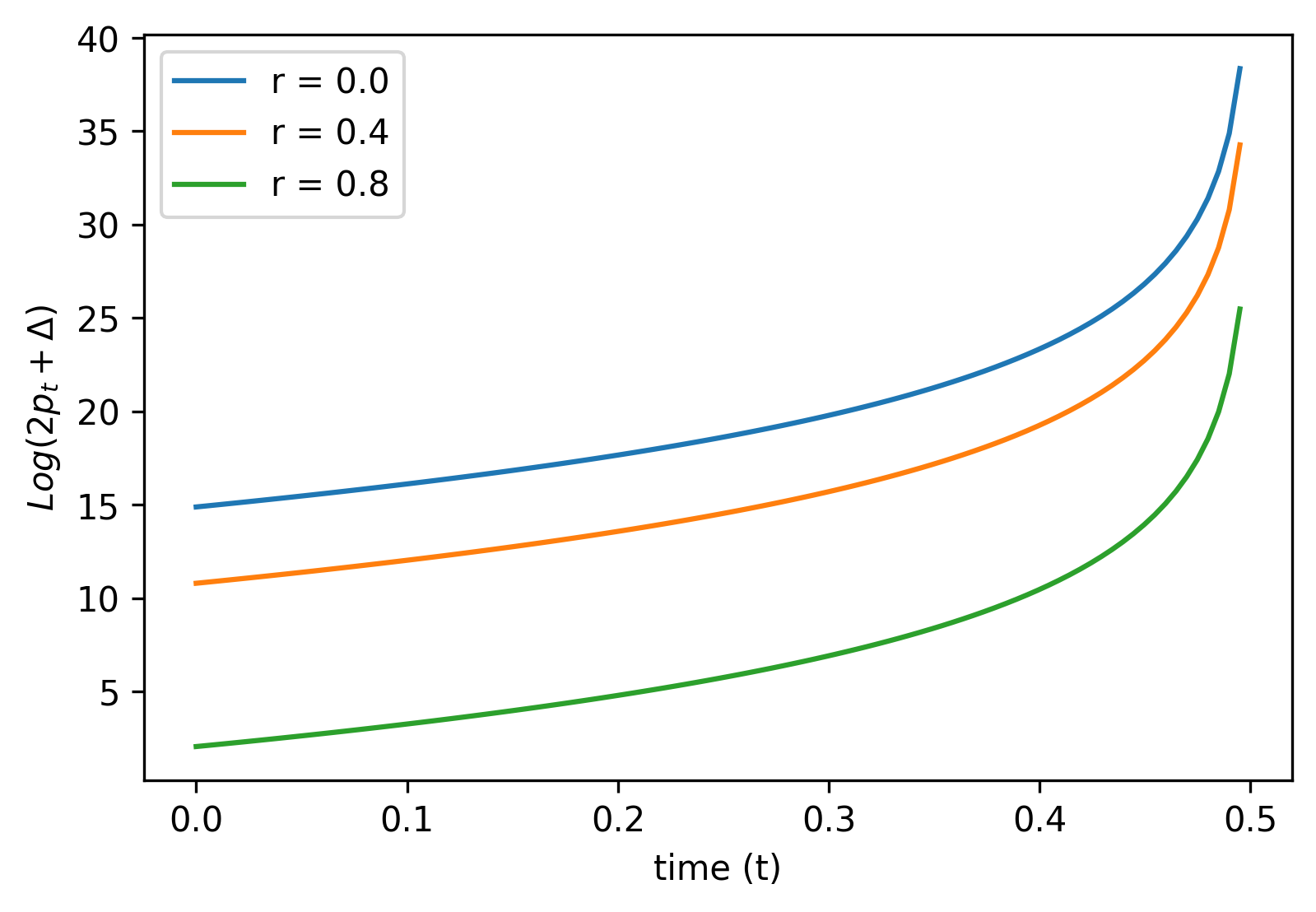}
\caption{Variation of $ Log(2p_t + \Delta) $ against time $ (t) $. }
\end{center}
\label{fig9}
\end{figure*}

In this section the author consider physical analysis of the model of gravitational collapse developed in this paper. It is found that for the value of $ b = 2 $ in equations (\ref{28}) to (\ref{31}), this model of gravitational collapse of compact star in $ f(R) = \xi R^4 $, obeys all the physical conditions from (\ref{32}) to (\ref{37}). For the value of $ \xi = 10^{4} $, all the physical quantities (\ref{28}) to (\ref{31}) and all conditions (\ref{32}) to (\ref{37}) have been plotted in figures (\ref{fig1}) to (\ref{fig9}) for different constant radius as a function of time. All the graphs are plotted as a logarithmic function due to convenience in analysis. From figures (\ref{fig1}) to (\ref{fig9}), it can be seen that all conditions are satisfied and thus this model is physically reliable. From figures (\ref{fig1}) to (\ref{fig9}) and also from equations (\ref{28}) to (\ref{31}) for $b = 2$, it can be clearly seen that gravitational collapse of the star ends in the singularity at time $t = 0.5$.  

\section{Physical properties of the stars undergoing gravitational collapse}\label{sec7}
The author has computed physical properties of the stars like energy density, heat flux density, and radial and tangential pressures undergoing gravitational collapse in section \ref{sec4}. However, other essential properties of the stars are governed by various parameters like four-acceleration and the expansion parameter of the fluid, the projection tensor, and the shearing tensors, which are helpful to investigate the inherent nature of gravitationally collapsing stars. \par The physical properties of the stable, compact stars have been investigated by various authors. Those studies found that the prior assumption of the isotropic nature of stable, compact stars holds. For example, authors have investigated the mass-radius relationship of neutron stars in $ f(R) $ gravity by specifying the equation of state relating the isotropic matter pressure and density.\cite{PhysRevD.93.023501}. The study helps investigate the stiffness of the equation of state to determine the maximum implementable value of central density of compact stars. Extreme neutron stars have also been investigated in the extended theory of gravity in the reference \cite{Astashenok_2015}. They utilized the isotropic behavior of the compact stars and discussed neutron stars with strong magnetic mean fields by solving Tolman-Oppenheimer-Volkoff equations in the extended theory of gravity. Also, recently authors have studied super-massive neutron stars and the causal limit of neutron star maximum mass in View of $GW190814$. They studied the properties of the neutron star by assuming the inherent isotropic nature of the star. The results obtained by them are found to be compatible with LIGO data. All these efforts show that the study of compact stars in $f(R)$ and extended theory of gravity can successfully be done by assuming the inherent isotropic nature of stable, compact stars. However, the study of gravitationally collapsing stars in $ f(R) $ gravity requires the model to be compatible with the anisotropic behavior of the stars. In this section, the author also shows that when the gravitational collapse of compact stars is governed by metric (\ref{5}) in $ f(R) $ gravity, the anisotropy of the gravitationally collapsing stars has to be taken into account.  
Firstly, the author computes various substantial quantities describing the physical properties of the stars undergoing gravitational collapse. The 4-velocity and the radial vector utilized in the equation (\ref{2}) for the space-time metric (\ref{5}) are given by

\begin{equation}
\label{38}
    u^{\alpha} = \left( \frac{1}{y(r)}, 0, 0, 0 \right)  \quad\mathrm{and}\quad v^{\alpha} = \left(0, \frac{1}{\sqrt{2}c(t)y'(r)}, 0, 0\right)
\end{equation}

The 4-acceleration of the fluid is computed from the following identity
\begin{equation}
\label{39}
    a^{\alpha} = u^{\alpha}_{\ ; \ \beta}u^{\beta} \quad\mathrm{and}\quad  a_{\alpha} = u_{\alpha ;  \beta}u^{\beta}
\end{equation}

which has only one non-zero radial component given by

\begin{equation}
\label{40}
    a^{1} = \frac{1}{2c(t)^2y'(r)y(r)}  \quad\mathrm{and}\quad  a_{1} = \frac{y'(r)}{y(r)}
\end{equation}

Now the author computes the expansion parameter as follow
\begin{equation}
\label{41}
   \Theta =  u^{\alpha}_{\ ; \ \alpha} = \frac{3 \dot{c}(t)}{c(t) y(r)}
\end{equation}

The shear tensor can be computed from the following identity
\begin{equation}
\label{42}
    \sigma_{\alpha \beta} = u_{(\alpha;\beta)} + a_{(\alpha}u_{\beta)} - \frac{1}{3}\Theta(g_{\alpha\beta} + u_{\alpha}u_{\beta})
\end{equation}

where $ h_{\alpha \beta} = g_{\alpha \beta} + u_{\alpha} u_{\beta} $ is known as projection tensor. components of shear tensor for the metric (\ref{5}) are computed as follow

\begin{equation}
\label{43}
    \sigma_{00} = \frac{\dot{c}}{c y}(-y^2 + y^2) = 0
\end{equation}

\begin{equation}
\label{44}
    \sigma_{01} = \sigma_{10} = \frac{y'}{2} - \frac{y'}{2} = 0
\end{equation}

\begin{equation}
\label{45}
    \sigma_{11} = \frac{2c\dot{c}y'^2}{y} - \frac{\dot{c}}{c y}(2 c^2 y'^2) = 0
\end{equation}

\begin{equation}
\label{46}
    \sigma_{22} = c \dot{c} y - \frac{\dot{c}}{c y}(c^2y^2) = 0
\end{equation}

\begin{equation}
\label{47}
    \sigma_{33} = (c \dot{c} y) sin^2\theta - \frac{\dot{c}}{c y}(c^2y^2)sin^2\theta = 0
\end{equation}

From equations (\ref{43}) to (\ref{47}), it can be seen that the components of shear tensor vanish for the interior metric (\ref{5}). All other components of the shear tensor are also zero because individual parts of the shear tensor (\ref{42}) vanishes identically. Thus, it can be concluded that the shear of gravitationally collapsing compact stars described by the metric (\ref{5}) is zero. Thus, the prior assumption of vanishing shear is completely justified for this study. However, it should be noted that the vanishing shear is only possible if gravitationally collapsing stars can be described by the separable form of the interior metric tensors. However, the phenomenon described by the non-separable form of metric tensor may be needed to include the study of shear and its evolution as described in the reference \cite{2008GReGr..40.2149P}.
Now the author computes the anisotropy $ (S) $ of the star undergoing the gravitational collapse for the $ f(R) = \xi R^4$ model under consideration. From equation (\ref{23}) and (\ref{24}),

\begin{equation}
\label{48}
    S = p_r - p_t = \frac{f_R'(y'^2+y y'' )}{2y y'^3 c^2} - \frac{f_R''}{2c^2y'^2}
\end{equation}

Now for the model $ f(R) = \xi R^4 $, the values of $ f_R' $ and $ f_R'' $ can be calculated by using equation (\ref{25}) as follow

\begin{equation}
\label{49}
    f_R' = -\frac{24 \xi R^3 y'}{y}
\end{equation}

\begin{equation}
\label{50}
    f_R'' = \frac{168 \xi R^3 y'^2}{y^2} - \frac{24 \xi R^3 y''}{y}
\end{equation}

Using equations (\ref{49}) and (\ref{50}) in equation (\ref{48}), the anisotropy is found to be

\begin{equation}
\label{51}
    S = p_r - p_t = -\frac{96\xi R^3}{c(t)^2y(r)^2}
\end{equation}

Equation (\ref{51}) implies that anisotropy vanishes for $ \xi = 0 $ only, in which case the solutions will reduce to GR only. Thus, we have to consider the anisotropic behavior of the compact stars to describe their gravitational collapse scenario using $ f(R) $ gravity. It also can be seen from equation (\ref{51}) that anisotropy does not vanish at the center of the gravitationally collapsing star. However, it is well-established that anisotropy must vanish at the center of stable, compact stars. Thus, it can be concluded that anisotropy is the inherent nature of gravitationally collapsing stars when described by the interior metric (\ref{5}) in $ f(R) $ gravity. 

\section{Junction conditions}
\label{sec8}
Junction conditions implied on the boundary surface of the star ensure that all physical parameters and quantities of the stars remain continuous across the hyper-surface. Thus, the matching junction condition across the collapsing time-like hyper-surface separating $ ds_-^2 $ and $ ds_+^2 $ are given by the following equations

\begin{equation}
\label{52}
    [g_{\alpha\beta}] = 0 \quad\mathrm{and}\quad   [K_{\alpha\beta}] = 0 
\end{equation}

\begin{equation}
\label{53}
    [R] = 0, \quad  n^{\alpha}[\partial_{\alpha}R] = 0 \quad\mathrm{and}\quad  n^{\alpha}[T_{\alpha\beta}] = 0 
\end{equation}

Where [] denotes the difference in quantity across the hyper-surface $ \Sigma $. Also $ n^{\alpha} $ is the radial vector. Now, the author has chosen the solutions of field equations (\ref{21}) to (\ref{24}) given in equations (\ref{26}) such that junction conditions mentioned in (\ref{53}) already satisfied as it can be seen from straight forward calculations also.\cite{Chowdhury:2019phb}. Junction conditions related to equation (\ref{52}) for the following general time dependant interior metric and exterior Vaidya metric are derived in the reference \cite{Chowdhury:2019phb}.
\begin{equation}
\label{54}
    ds_-^2 = -e^{2\nu(t,r)}dt^2 + e^{2\psi(t,r)}dr^2 + Q^2(t,r)(d\theta^2 + sin^2\theta d\phi^2)
\end{equation}

\begin{equation}
\label{55}
    ds_+^2 = -\left( 1 - \frac{2m(u)}{\Tilde{r}} \right)du^2 -2 du d\Tilde{r} + \Tilde{r}^2(d\theta^2 + sin^2\theta d\phi^2)
\end{equation}

The first junction condition corresponding to continuity of metric tensor gives the following identity

\begin{equation}
\label{56}
    M_{\Sigma} = \frac{Q}{2}\left[ 1 - e^{-2\psi}\left( \frac{dQ}{dr} \right)^2 +  e^{-2\nu}\left( \frac{dQ}{dt} \right)^2  \right]_{\Sigma}
\end{equation}

Where the $M_{\Sigma}$ is the Misner-Sharp mass corresponding to the total mass inside the hyper-surface $ \Sigma $. Now the second junction condition corresponding to the continuity of extrinsic curvature $ K_{\alpha \beta} $ is given by

\begin{multline}
\label{57}
    \frac{Q}{2}e^{-(\nu+\psi)}\left( 2\frac{\dot{Q}'}{Q} - 2\frac{\dot{Q}}{Q}\frac{\dot{\psi}}{\psi} - 2\frac{\nu'}{\nu}\frac{\dot{Q}}{Q}\right) \\+
    \frac{Q}{2} e^{-2\nu}\left( 2\frac{\ddot{Q}}{Q} - 2\frac{\dot{Q}}{Q}\frac{\dot{\nu}}{\nu} + \frac{e^{2\nu}}{Q^2} + \frac{\dot{Q}^2}{Q^2} - e^{2(\nu-\psi)}\left( \frac{Q'^2}{Q^2} - 2\frac{\nu'}{\nu}\frac{Q'}{Q} \right) \right) \bigg |_{\Sigma} = 0
\end{multline}
Now by comparing the interior metric (\ref{54}) with metric (\ref{5}), we get $ e^{v(t,r)} = y(r) $, $ e^{\psi (t,r)} = \sqrt{2}c(t)y'(r) $ and $ Q(t,r) = c(t) y(r) $. Substituting these values in equations (\ref{56}) and (\ref{57}), we get the following junction conditions for metric (\ref{5})

\begin{equation}
\label{58}
 M_{\Sigma} =  \frac{c y}{2}\left[ \frac{1}{2} + \dot{c}^2 \right]_{\Sigma}
\end{equation}

\begin{multline}
\label{59}
\frac{1}{2\sqrt{2}y'}\left(\frac{2\dot{c}y'}{c y} - 2\frac{\dot{c}^2}{c^2}\frac{1}{log(\sqrt{2}c y')} - \frac{2\dot{c}}{c y log(y)} + \frac{2\sqrt{2}\ddot{c}y'}{y}\right) \\+ \frac{1}{2\sqrt{2}y'}\left( \frac{\sqrt{2}y'}{c y} + \frac{\sqrt{2}y'\dot{c}^2}{c y} - \frac{y'}{\sqrt{2}c y} + \frac{\sqrt{2}}{c y log(y)} \right)\bigg |_{\Sigma} = 0    
\end{multline}

Where $ \Sigma $ in the equations (\ref{58}) and (\ref{59}) is the matching hyper-surface between the interior metric (\ref{5}) and outer Vaidya metric (\ref{55}) and  $M_{\Sigma}$ is corresponding Misner-Sharp mass within the hyper-surface $ \Sigma $. Thus, equations (\ref{58}) and (\ref{59}) are the junction conditions corresponding to the metric (\ref{5}) under study. 

\section{Conclusion}
\label{sec9}
In this paper, the author obtained analytical solutions for the gravitational collapse of the compact stars in $ f(R) = \xi R^4 $ gravity. Where $ f(R) = \xi R^4 $ is the physically important model of $ f(R) $ gravity that describes the existence of the slow-roll inflation. By reviewing the gravitational collapse in $ f(R) $ gravity, the correct form of field equations has been re-derived for the interior metric (\ref{5}). From that equations, analytical solutions for the gravitational collapse in $ f(R) = \xi R^4 $ gravity have been obtained. By imposing various physical conditions on the model, the value of model parameter $ b $ has been found such that the model becomes physically acceptable. By plotting graphs of various conditions on the model, it is shown that the model obeys all the conditions, and thus it is physically acceptable. In physical analysis section it has been shown that this model describes gravitational collapse of star starting from $t = 0$ to $ t = 0.5 $ in $ f(R) = \xi R^4 $ gravity. At $ t = 0.5 $, the gravitational collapse of the star ends in the singularity. The author also calculated and studied various physical parameters and quantities of the star undergoing gravitational collapse and imposed the junction conditions so that model became physically well behaved at the surface of the gravitationally collapsing star. Thus, the model of gravitational collapse of compact stars presented in this paper is physically acceptable, completely well behaved and all the physical properties of the star undergoing gravitational collapse is thoroughly studied in this paper.

\bibliographystyle{unsrt}
\bibliography{main.bib}

\end{document}